\documentclass[prd,showpacs,11pt]{revtex4}
\usepackage{amsmath}
\usepackage{amssymb}
\usepackage[dvips]{graphicx}

\begin{document}
\title{Reconstructing $f(R,T)$ gravity from holographic dark energy}

\author{M.~J.~S.~Houndjo}
\email{sthoundjo@yahoo.fr}
\affiliation{ICRA - Centro Brasileiro de Pesquisas F\'{i}sicas - CBPF - Rua Dr. Xavier Sigaud, 150, Urca, 22290-180, Rio de Janeiro, Brazil}
\author{Oliver F. Piattella}
\email{oliver.piattella@gmail.com}
\affiliation{Department of Physics, Universidade Federal do Esp\'irito Santo, avenida Ferrari 514, 29075-910 Vit\'oria, Esp\'irito Santo, Brazil}

%%%%%%%%%%%%%%%%%%%%%%%%%%%%%%%%%%%%%%%%%%%%%%%%%%%%%%%%%%%%%%%%%%%%%%%%%%%%%%%%%%%%%%%%%%%%%%%%%%%%%%%%%%%%%%%%%%%%%

\begin{abstract}
\textbf{We consider cosmological scenarios based on $f(R,T)$ theories of gravity ($R$ is the Ricci scalar and $T$ is the trace of the energy-momentum tensor) and numerically reconstruct the function $f(R,T)$ which is able to reproduce the same expansion history generated, in the standard General Relativity theory, by dark matter and holographic dark energy. We consider two special $f(R,T)$ models: in the first instance, we investigate the modification $R + 2f(T)$, i.e. the usual Einstein-Hilbert term plus a $f(T)$ correction. In the second instance, we consider a $f(R)+\lambda T$ theory, i.e. a $T$ correction to the renown $f(R)$ theory of gravity.}
\end{abstract}


\maketitle

%%%%%%%%%%%%%%%%%%%%%%%%%%%%%%%%%%%%%%%%%%%%%%%%%%%%%%%%%%%%%%%%%%%%%%%%%%%%%%%%%%%%%%%%%%%%%%%%%%%%%%%%%%%%%%%

\section{Introduction}

The nature of dark matter and dark energy is one of the most important issues today in physics. There are strong observational evidences in astrophysics and cosmology for the existence of these two components of the cosmic energy budget, indicating that about $95\%$ of the Universe is composed by dark matter (about $25\%$) and by dark energy (about $70\%$), but no direct detection has been reported until now. The usual candidates \textbf{for} dark matter (neutralinos and axions, for example) and dark energy (cosmological constant, quintessence, etc.) lead to very robust scenarios, but at same time they must face theoretical and observational issues. For recent reviews on the subject, see for example \cite{Padmanabhan:2002ji, Peebles:2002gy, Sahni:2004ai, Bertone:2004pz, Sahni:2006pa, Copeland:2006wr, Frieman:2008sn, Martin:2008qp, Caldwell:2009ix, Li:2011sd}.

The strongest issue is perhaps the one regarding dark energy as the vacuum expectation value of some quantum field, which would be a natural candidate, but whose correct theoretical value could be predicted only in the framework of a complete theory of quantum gravity, which still we do not possess. Nevertheless, it is possible, at least, to guess some of the features of this theory. In particular, the holographic principle \cite{'tHooft:1993gx, Susskind:1994vu, Bousso:2002ju} may shed some light on the dark energy problem. According to this principle, in presence of gravity the number of the degrees of freedom of a local quantum system would be related to the area of its boundary, rather than to the volume of the system (as expected when gravity is absent). Following this \textbf{idea}, in \cite{Cohen:1998zx} the authors suggested an entanglement relation between the infrared and ultraviolet \textbf{cutoffs} due to the limitation set by the formation of a black hole, which sets an upper bound for the vacuum energy. We can then interpret the ultraviolet cutoff as the vacuum density value, but still we need an ansatz for the infrared cutoff. As a candidate for such distance, in \cite{Li:2004rb, Huang:2004ai} the authors propose and investigate the future event horizon, tested against type Ia supernovae data and cosmic microwave background anisotropies in \cite{Zhang:2005hs, Zhang:2007sh}. We shall present more detail on this in Sec.~\ref{Sec:HolDE}. 

Adding new components of dark energy to the whole energy budget in order to explain the current observation is a way, but not the only one. Since General Relativity has been thoroughly tested up to solar system scales, it may be possible that the Einstein-Hilbert action contain corrections on larger, cosmological, scales thereby candidating as possible explanation of the evolution of the universe. Such modifications should be, in principle, composed by higher order curvature invariant terms (such as $R^2$, $R_{\mu\nu}R^{\mu\nu}$, etc) but also by non-trivial coupling between matter or fields and geometry. See for example \cite{Nojiri:2006ri, Nojiri:2010wj, Amendola:2006kh, Amendola:2006we, Capozziello:2007ec, Sotiriou:2008rp, DeFelice:2010aj} for some reviews on the subject (especially on $f(R)$ theory). It is also worth pointing out that these terms should naturally emerge as quantum corrections in the low energy effective action of quantum gravity or string theory \cite{Buchbinder:1992rb}.

In this paper we connect these two approaches, considering a $f(R,T)$ theory of gravity, where $R$ is the Ricci scalar, whereas $T$ is the trace of the stress-energy momentum. This modified gravity theory has been recently introduced in \cite{Harko:2011kv}, where the authors derived the field equations and considered several cases, relevant in cosmology and astrophysics. As for the former, $f(R,T)$ models have been constructed describing the transition from the matter dominated phase to the late times accelerated one \cite{Houndjo:2011tu}. 

Our task here, is to find out which form the function $f(R,T)$ has to have in order to reproduce the same properties of the holographic dark energy proposed in \cite{Li:2004rb}. To this purpose, we employ the same reconstruction scheme proposed and employed in \cite{Capozziello:2005ku, Nojiri:2006gh, Nojiri:2006jy, Nojiri:2006be, Wu:2007tn}. For reference, in order to track the contribution of the $T$ part of the action in the reconstruction, we consider two special $f(R,T)$ models: in the first instance, we investigate the modification $R + 2f(T)$, i.e. the usual Einstein-Hilbert term plus a $f(T)$ correction. In the second instance we consider a $f(R)+\lambda T$ theory, i.e. a $T$ correction to the renown $f(R)$ gravity. In both cases, we consider dark energy accompanied by a pressureless matter component (which would determine $T$).

The paper is organised as follows. In Sec.~\ref{Sec:HolDE}, the equations of motion are established and the holographic dark energy introduced. In Sec.~\ref{Sec:Simpl} and \ref{Sec:ComplCase} the above mentioned cases are analysed. Finally, Sec.~\ref{Sec:DiscConcl} is devoted to discussion and conclusions.

We use $8\pi G = c = 1$ units and adopt the metric formalism, i.e. \textbf{the variation of the action is considered with respect to the metric quantities.}

%%%%%%%%%%%%%%%%%%%%%%%%%%%%%%%%%%%%%%%%%%%%%%%%%%%%%%%%%%%%%%%%%%%%%%%%%%%%%%%%%%%%%%%%%%%%%%%%%%%%%%%%%%%%%%%%

\section{$f(R,T)$ gravity and holographic Dark Energy}\label{Sec:HolDE}

In \cite{Harko:2011kv}, the following modification of \textbf{Einstein's} theory is proposed:
\begin{equation}\label{actionfRT}
 S = \frac{1}{2}\int f(R,T) \sqrt{-g}\;d^4x + \int L_{\rm m} \sqrt{-g}\;d^4x\;,
\end{equation}
where $f(R,T)$ is an arbitrary function of the Ricci scalar $R$ and of the trace $T$ of the energy-momentum tensor, defined as
\begin{equation}
 T_{\mu\nu} = -\frac{2}{\sqrt{-g}}\frac{\delta\left(\sqrt{-g}L_{\rm m}\right)}{\delta g^{\mu\nu}}\;,
\end{equation}
where $L_{\rm m}$ is the matter Lagrangian density. We assume the matter lagrangian to depend on the metric, so that
\begin{equation}
 T_{\mu\nu} = g_{\mu\nu}L_{\rm m} - 2\frac{\partial L_{\rm m}}{\partial g^{\mu\nu}}\;.
\end{equation}
Varying action \eqref{actionfRT} with respect to the metric $g^{\mu\nu}$, one obtains \cite{Harko:2011kv}
\begin{equation}\label{Eqmod}
 f_R(R,T) R_{\mu\nu} - \frac{1}{2}f(R,T)g_{\mu\nu} + \left(g_{\mu\nu}\square - \nabla_\mu\nabla_\nu\right)f_R(R,T) = T_{\mu\nu} - f_T(R,T)T_{\mu\nu} - f_T(R,T)\Theta_{\mu\nu}\;,
\end{equation}
where the \textbf{subscripts} $R$ or $T$ \textbf{imply} derivation with respect that quantity and we have also defined
\begin{equation}
 \Theta_{\mu\nu} \equiv g^{\alpha\beta}\frac{\delta T_{\alpha\beta}}{\delta g^{\mu\nu}}\;.
\end{equation}
\textbf{Planning} a cosmological application, we assume matter to be described by a perfect fluid energy-momentum tensor
\begin{equation}
 T_{\mu\nu} = \left(\rho + p\right)u_\mu u_\nu - p g_{\mu\nu}\;,
\end{equation}
and that $L_{\rm m} = -p$, so that we have 
\begin{equation}
 \Theta_{\mu\nu} = -2T_{\mu\nu} - p g_{\mu\nu}\;,
\end{equation}
and Eq.~\eqref{Eqmod} simplifies as
\begin{equation}\label{Eqmod2}
 f_R(R,T) R_{\mu\nu} - \frac{1}{2}f(R,T)g_{\mu\nu} + \left(g_{\mu\nu}\square - \nabla_\mu\nabla_\nu\right)f_R(R,T) = T_{\mu\nu} + f_T(R,T)T_{\mu\nu} + p f_T(R,T) g_{\mu\nu}\;.
\end{equation}
In order to compare \textbf{it} with \textbf{Einstein's}, we cast the above equation as follows:
\begin{eqnarray}\label{Eqmod3}
 G_{\mu\nu} &=& \frac{1 + f_T(R,T)}{f_R(R,T)}T_{\mu\nu} + \frac{1}{f_R(R,T)}p f_T(R,T) g_{\mu\nu} - \frac{1}{f_R(R,T)}\left(g_{\mu\nu}\square - \nabla_\mu\nabla_\nu\right)f_R(R,T)\nonumber\\ &+& \frac{1}{2f_R(R,T)}f(R,T)g_{\mu\nu} - \frac{1}{2}g_{\mu\nu} R\;,
\end{eqnarray}
where $G_{\mu\nu} \equiv R_{\mu\nu} - Rg_{\mu\nu}/2$ is the Einstein tensor. Now we can identify
\begin{equation}\label{Effmatt}
 \tilde{T}_{\mu\nu}^{(m)} = \frac{1 + f_T(R,T)}{f_R(R,T)}T_{\mu\nu} + \frac{1}{f_R(R,T)}p f_T(R,T) g_{\mu\nu}\;,
\end{equation}
as the \textit{effective} matter energy-momentum tensor and
\begin{equation}\label{Effgeom}
 \tilde{T}_{\mu\nu}^{(geom)} = - \frac{1}{f_R(R,T)}\left(g_{\mu\nu}\square - \nabla_\mu\nabla_\nu\right)f_R(R,T) + \frac{1}{2f_R(R,T)}f(R,T)g_{\mu\nu} - \frac{1}{2}g_{\mu\nu} R\;,
\end{equation}
as the energy-momentum tensor of a ``geometric'' matter component. 

We now assume a \textbf{background} described by the Friedmann-Lema\^{\i}tre-Robertson-Walker metric
\begin{equation}\label{RWmet}
 ds^2 = dt^2 - a(t)^2\delta_{ij}dx^idx^j\;, 
\end{equation}
with spatially flat hypersurfaces, and find a form for the function $f(R,T)$ which is able to reconstruct \textbf{holographic dark energy}. 

\subsection{Holographic Dark Energy}

According to the holographic principle \cite{'tHooft:1993gx, Susskind:1994vu, Bousso:2002ju} an entanglement relation between the infrared (IR) and ultraviolet (UV) cut-offs of a quantum theory, due to the limitation set by the formation of a black hole, sets an upper bound for the vacuum energy \cite{Cohen:1998zx}:
\begin{equation}\label{vacen}
 \rho_{\rm v} = \frac{3b^2}{L^2}\;,
\end{equation}
where $b$ is a free parameter and the IR (large scales) cutoff $L$ needs to be specified by an ansatz. We are interested in the one proposed in \cite{Li:2004rb}:
\begin{equation}\label{anshol}
 L = R_{\rm h} = a\int_t^\infty \frac{dt'}{a(t')} = a\int_a^\infty \frac{d\bar{a}}{H(\bar{a})\bar{a}^{2}}\;,
\end{equation}
i.e. the future event horizon, that is the distance covered by a photon from now until the remote future. Note that the very presence of a vacuum energy component makes the above integration finite. We consider a model composed by holographic dark energy plus ordinary pressureless matter, i.e.
\begin{equation}\label{Friede}
 3H^2 = \rho_{\rm v} + \rho_{\rm m} = \rho_{\rm v} + \rho_{\rm m0}(1 + z)^3\;,
\end{equation}
where $H \equiv \dot{a}/a$ is the Hubble parameter and the dot denotes derivation with respect to the cosmic time. Introduce the critical energy density $\rho_{\rm cr} := 3H^2$\textbf{, we define}
\begin{equation}\label{oliver15}
\Omega_{\rm v} := \frac{\rho_{\rm v}}{\rho_{\rm cr}}=\frac{b^2}{R^2_{\rm h}H^2}\;,
\end{equation}
Using Eqs.\eqref{vacen} and \eqref{anshol}, it is easy to show that 
\begin{equation}\label{oliver16}
\dot{R}_{\rm h} = \frac{b}{\sqrt{\Omega_{\rm v}}} - 1\;.
\end{equation}
The holographic dark energy density $\rho_{\rm v}$ evolves according to the conservation law
\begin{equation}\label{oliver17}
\dot{\rho}_{\rm v} + 3H\rho_{\rm v}\left(1 + w_{\rm v}\right) = 0\;,
\end{equation}
because in Eq.~\eqref{Friede} we have implicitly assumed the matter component to conserve separately. Now, using Eqs.~\eqref{vacen}, \eqref{anshol} and \eqref{oliver16}, one can find
\begin{equation}\label{oliver18}
\dot{\rho}_{\rm v} = -\frac{2}{R_{\rm h}}\left(\frac{b}{\sqrt{\Omega_{\rm v}}}-1\right)\rho_{\rm v}\,.
\end{equation}
Comparing \eqref{oliver18} with \eqref{oliver17} one can read off
\begin{equation}\label{oliver19}
w_{\rm v} = -\left(\frac{1}{3} + \frac{2\sqrt{\Omega_{\rm v}}}{3b}\right)\,\,.
\end{equation}
Moreover, combining Eq.~\eqref{anshol} with Eqs.~\eqref{vacen} and \eqref{Friede}, the evolution for $\Omega_{\rm v}$ is determined by the following equation:
\begin{equation}\label{OmegavEvo}
 \Omega_{\rm v}' = -\left(1 + \frac{2\sqrt{\Omega_{\rm v}}}{b}\right)\frac{1}{1 + z}\Omega_{\rm v}\left(1 - \Omega_{\rm v}\right)\;,
\end{equation}
where the prime denotes derivation with respect to the redshift $z$. Testing this model against type Ia supernovae and cosmic microwave background anisotropies, $b$ turns out to be constrained around unity, with the case $b < 1$ favoured \cite{Zhang:2005hs, Zhang:2007sh}. Note that, from Eq.~\eqref{oliver19}, $b < 1$ means that the universe will end up in a phantom phase. For more comprehensive analysis of holographic dark energy models, we refer the reader to \cite{Pavon:2005yx, delCampo:2011jp}.

In the next section we investigate a simple case of reconstruction of $f(R,T)$.

\section{A simple case}\label{Sec:Simpl}

We now consider a single perfect fluid model with density $\rho$ and pressure $p$, together with the following ansatz (one of the first considered in \cite{Harko:2011kv}):
\begin{equation}
 f(R,T) = R + 2f(T)\;,
\end{equation}
i.e. the action is given by the same Einstein-Hilbert one plus a function of $T$. This is a particularly interesting choice since, from Eqs.~\eqref{Effmatt} and \eqref{Effgeom}, we get
\begin{equation}\label{Tmtil}
 \tilde{T}_{\mu\nu} = \left(1 + 2f_T\right)T_{\mu\nu} + 2p f_T g_{\mu\nu} + f(T)g_{\mu\nu}\;.
\end{equation}
For $p = 0$ one has $T = \rho$ and, choosing $f(T) = \lambda T$ one can construct a model with an effective cosmological constant \cite{Poplawski:2006ey}. From Eq.~\eqref{Tmtil} one can read off the effective energy density and pressure of the universe content:
\begin{eqnarray}
\label{rhotot} 3H^2 &=& \rho_{\rm eff} = \left(1 + 2f_T\right)\rho + 2p f_T + f(T)\;,\\
\label{ptot} -2\dot{H} - 3H^2 &=& p_{\rm eff} = p - f(T)\;,
\end{eqnarray}
and therefore a dark energy component may appear, even if we are considering a single perfect fluid model. From Eqs.~\eqref{rhotot} and \eqref{ptot}, it is clear that we can pick out a ``fictitious'' component, due to $f(T)$, described by
\begin{eqnarray}
 \rho_{f} &=& 2f_T\rho + 2p f_T + f(T)\;,\\
 p_{f} &=& - f(T)\;,
\end{eqnarray}
and, provided $f$ positive, it may well describe a dark energy component, since its pressure is negative. In order to reconstruct the function $f$ starting from the holographic principle, we note that the equation of state parameter of the dark component induced by $f$ is
\begin{equation}
 w_{f} = -\frac{f(T)}{2(\rho + p)f_T + f(T)}\;,
\end{equation}
and we identify it with $w_{\rm v}$, the one provided by the holographic dark energy in Eq.~\eqref{oliver19}:
\begin{equation}\label{EqfT}
 \frac{f(T)}{2(\rho + p)f_T + f(T)} = \frac{1}{3} + \frac{2\sqrt{\Omega_{\rm v}}}{3b}\;.
\end{equation}
For the standard model given in Eq.~\eqref{Friede}, consider the fluid component to be pressureless matter, i.e. $p = 0$. We are left to solve the following system of equations:
\begin{eqnarray}
\label{fevo} \rho\frac{df(\rho)}{d\rho} &=& f(\rho)\frac{b - \sqrt{\Omega_{\rm v}}}{b + 2\sqrt{\Omega_{\rm v}}}\;,\\
\label{Omevo} \frac{d\Omega_{\rm v}}{d\rho} &=& -\frac{1}{3\rho}\left(1 + \frac{2\sqrt{\Omega_{\rm v}}}{b}\right)\Omega_{\rm v}\left(1 - \Omega_{\rm v}\right)\;,
\end{eqnarray}
where we have used $T = \rho$, because we are considering pressureless matter. In order to solve the above system, we have to fix some initial conditions. Clearly, $\Omega_{\rm v}(\rho = \rho_0) = 1 - \Omega_{\rm m0}$, and we choose $\Omega_{\rm m0} = 0.3$, accordingly with current cosmological observation. For the initial condition on $f$, from Eq.~\eqref{rhotot} (with $p = 0$) we write
\begin{equation}\label{friedeqf0}
 \left[1 + 2f_{T}(\rho_0)\right]\Omega_{\rm m0} + \frac{f(\rho_0)}{3H_0^2} = 1\;.
\end{equation}
Evaluating Eq.~\eqref{fevo} today and combining it with Eq.~\eqref{friedeqf0} we find the following algebraic equation determining the initial condition on $f$:
\begin{equation}\label{f0cond}
 f(\rho_0)\left(2\frac{b - \sqrt{\Omega_{\rm v0}}}{b + 2\sqrt{\Omega_{\rm v0}}} + 1\right) = 3H_0^2\Omega_{\rm v0}\;.
\end{equation}
As we expected, when $\Omega_{\rm v0} = 0$, then $f(\rho_0) = 0$ and Eq.~\eqref{fevo} implies that $f$ is identically vanishing.\\
\begin{figure}[htbp]
  \includegraphics[width=0.45\columnwidth]{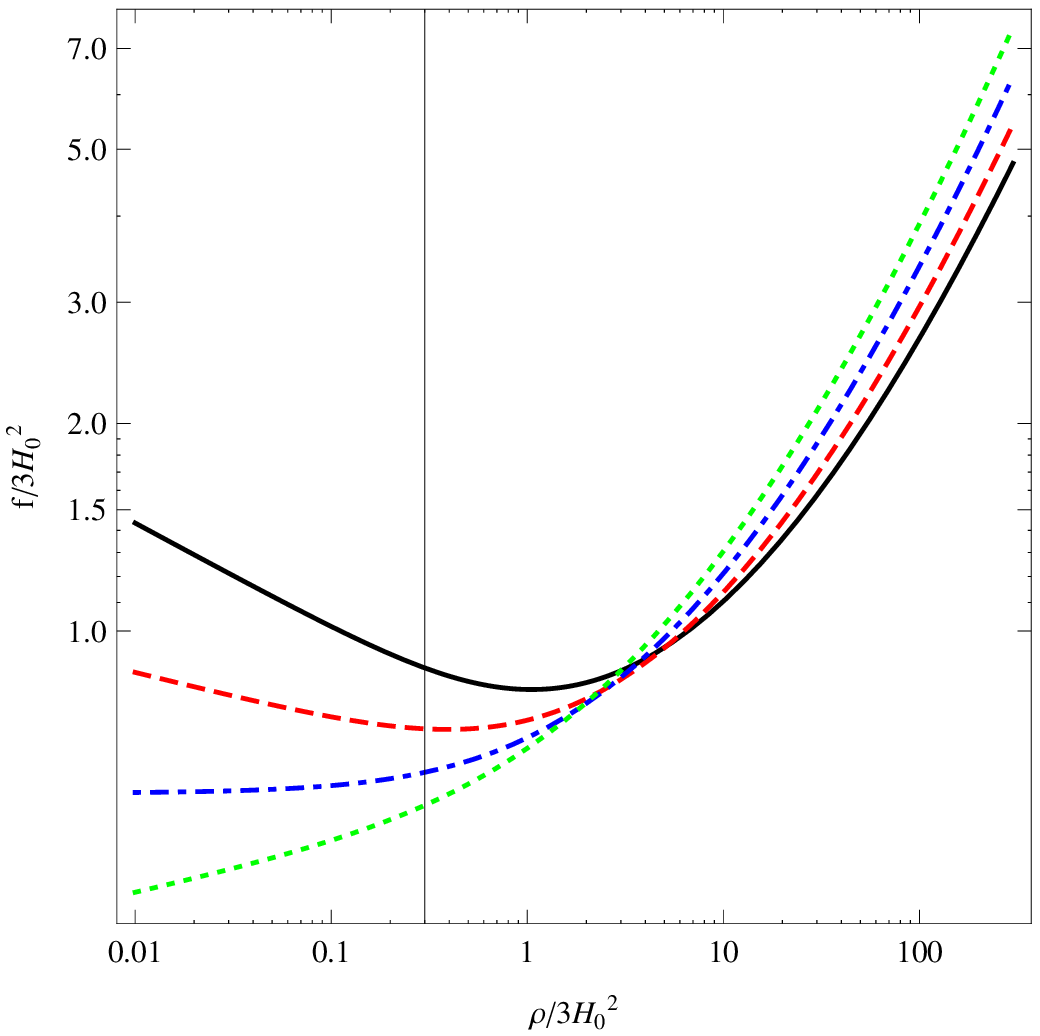}\;\includegraphics[width=0.45\columnwidth]{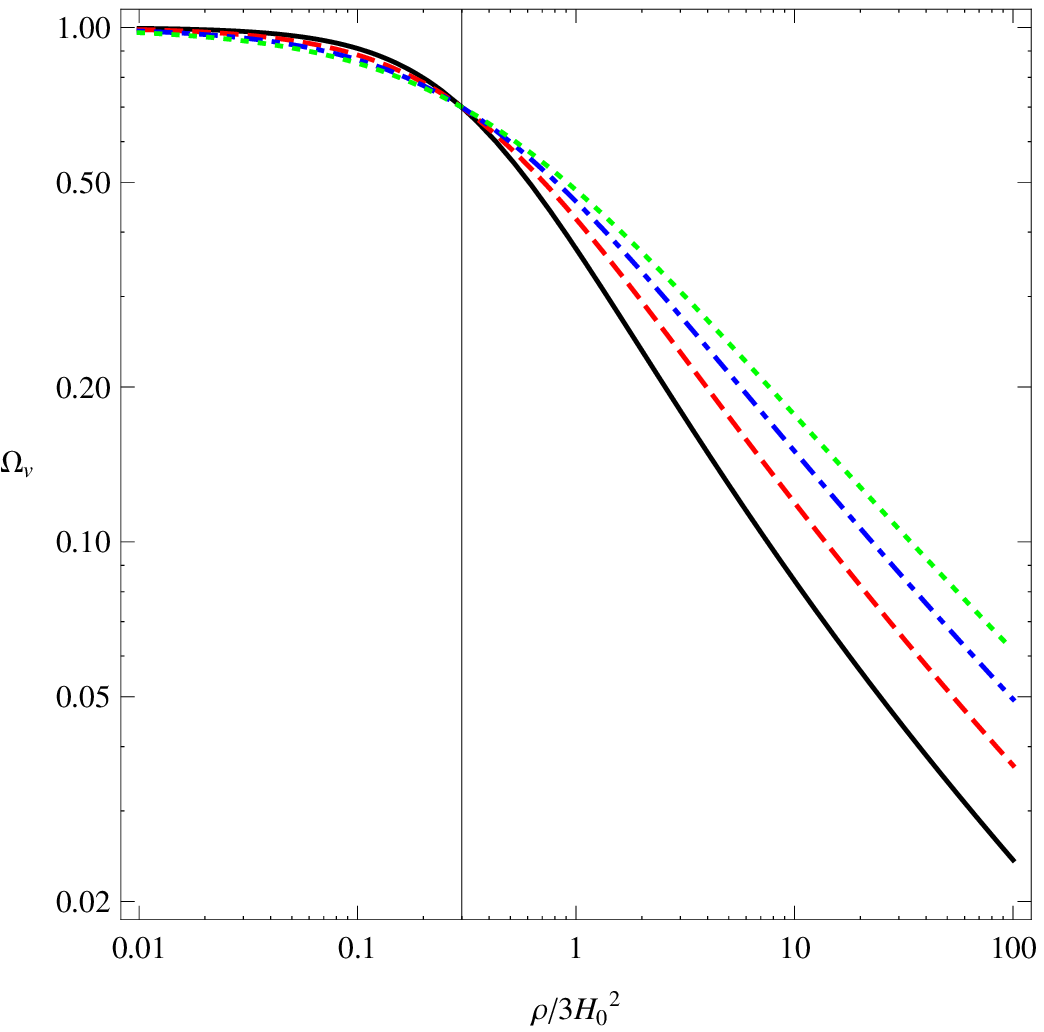}
\caption{Left panel: evolution of $f(\rho)$. Right panel: evolution of $\Omega_{\rm v}$. The cases considered are $b = 0.6, 0.8, 1.0, 1.2$ (solid black, dashed red, dot-dashed blue and dotted green, respectively). We have chosen as initial conditions in $\Omega_{\rm m0} = 0.3$ the values $\Omega_{\rm v0} = 1 - \Omega_{\rm m0} = 0.7$ and $f_0$ given by Eq.~\eqref{f0cond}. Note that $f$ and $\rho$ are normalised to $3H_0^2$. The vertical lines in the plots represent $\rho = \rho_0$, i.e. the present instant.}
\label{Fig1}
\end{figure}\\
In \figurename{ \ref{Fig1}} we plot the solution of the system of differential equations \eqref{fevo} and \eqref{Omevo}. Note that we normalise $f$ and $\rho$ to $3H_0^2$. As expected from Eq.~\eqref{fevo}, for large values of $\rho$, i.e. far in the past, $f \propto \rho$ because the dark energy component is subdominant. The actual difference among the various choices of $b$ takes place at late times, for small values of $\rho$. Again from inspection of Eq.~\eqref{fevo}, we can see that for large values of $b$ the linear evolution $f \propto \rho$ is again solution. That is why in the left panel of \figurename{ \ref{Fig1}} the curve seems to ``straighten up'' for increasing $b$. 

\begin{figure}[htbp]
  \includegraphics[width=0.45\columnwidth]{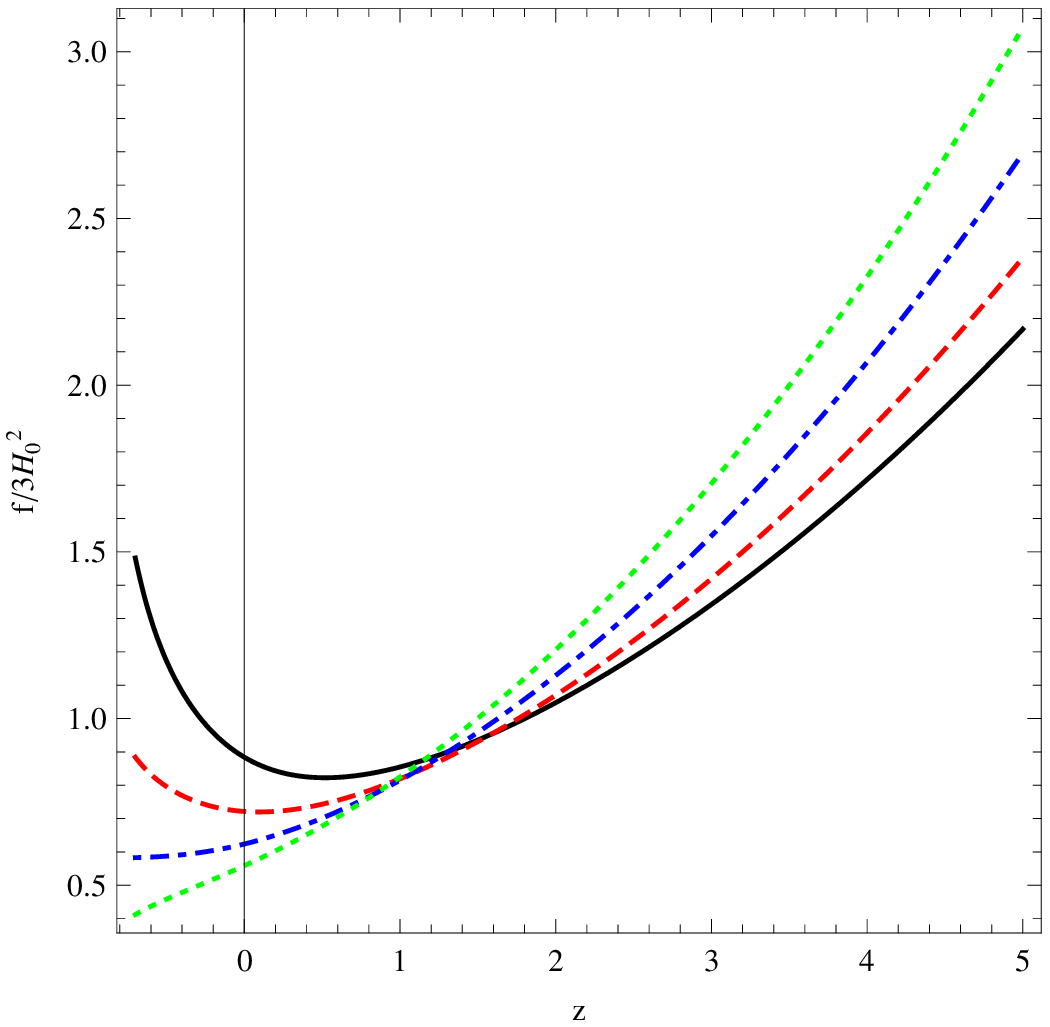}\;\includegraphics[width=0.45\columnwidth]{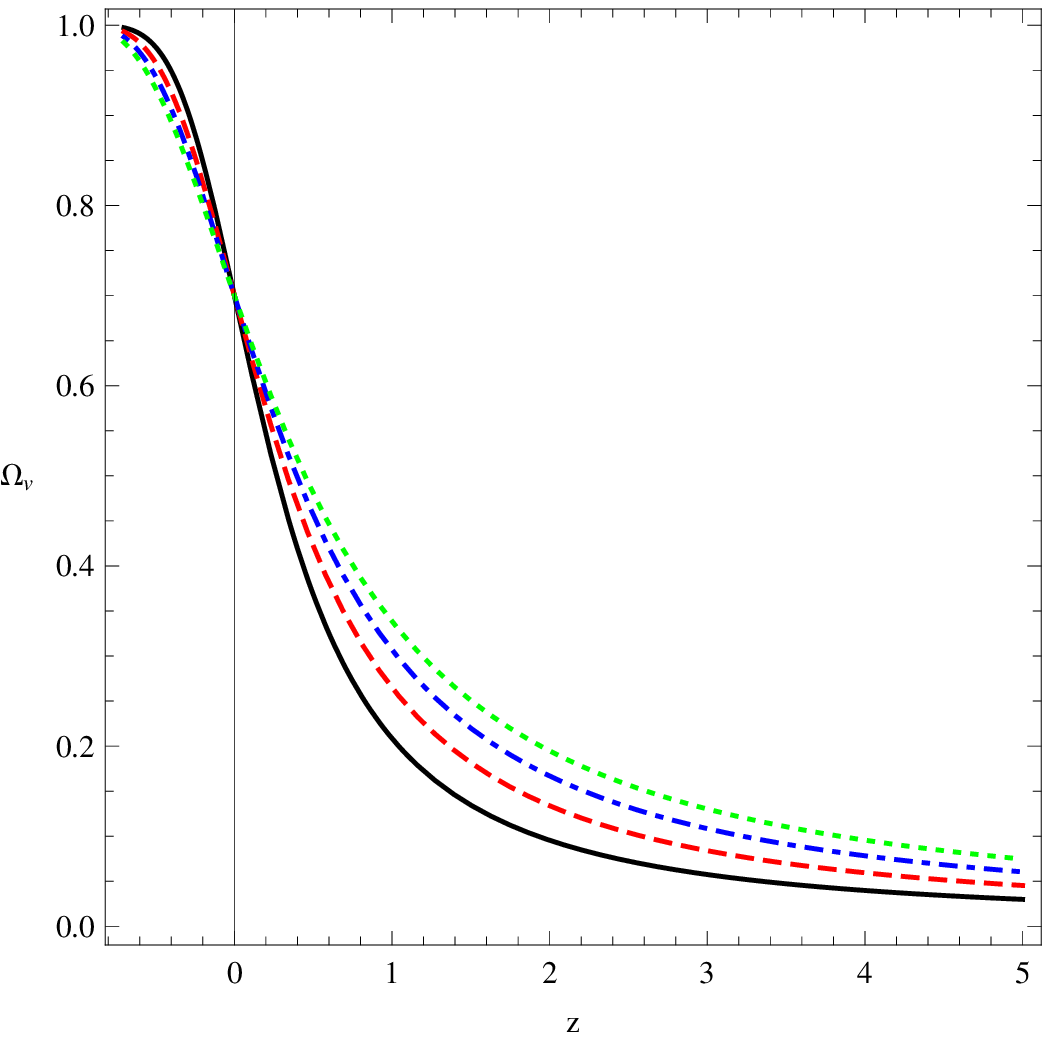}
\caption{Same as \figurename{ \ref{Fig1}}, but with the redshift $z$ as independent variable.}
\label{Fig2}
\end{figure}

In \figurename{ \ref{Fig2}} we plot the same quantities, but as functions of the redshift, in order to make clearer their cosmological evolution.

A final remark about the future evolution. It is clear from Eq.~\eqref{fevo}, that when $\Omega_{\rm v} \to 1$, in the remote future, the solution for $f$ gets the asymptotic form
\begin{equation}
 f(\rho) \propto \rho^{\frac{b - 1}{b + 2}}\;.
\end{equation}
Therefore, we would have a future singularity for $-2 < b < 1$, as it appears for the relevant cases of \figurename{ \ref{Fig1}} and \figurename{ \ref{Fig2}}. The special case $b = 1$ implies an asymptotically constant $f$.

We now turn our discussion on a more general case, where the curvature $R$ \textbf{comes into} the action \textbf{as} a function to be determined.

%%%%%%%%%%%%%%%%%%%%%%%%%%%%%%%%%%%%%%%%%%%%%%%%%%%%%%%%%%%%%%%%%%%%%%%%%%%%%%%%%%%%%%%%%%%%%%%%%%%%%%%%%%%%%%%%%%

\section{A more complicated case}\label{Sec:ComplCase}

Now we turn our attention to the special case
\begin{equation}\label{complcase}
 f(R,T) = f(R) + \lambda T\;,
\end{equation}
i.e. a $T$-linear correction to the class of $f(R)$ theories. With the ansatz \eqref{complcase} the matter content \eqref{Effmatt} is ``corrected'' as follows:
\begin{equation}\label{Effmattcomplcase}
 \tilde{T}_{\mu\nu}^{(m)} = \frac{1 + \lambda}{f_R(R)}T_{\mu\nu} + \frac{1}{f_R(R)}\lambda p g_{\mu\nu}\;,
\end{equation}
whereas the geometry induced stress-energy tensor is
\begin{equation}\label{Effgeomcomplcase}
 \tilde{T}_{\mu\nu}^{(geom)} = - \frac{1}{f_R(R)}\left(g_{\mu\nu}\square - \nabla_\mu\nabla_\nu\right)f_R(R) + \frac{1}{2f_R(R)}[f(R) + \lambda T]g_{\mu\nu} - \frac{1}{2}g_{\mu\nu} R\;.
\end{equation}
Our aim is now to reconstruct the form of the $f(R)$ which is able to reproduce the holographic dark energy paradigm. We again consider a pressureless perfect fluid with density $\rho$ and again assume metric \eqref{RWmet}. From Eqs.~\eqref{Effmattcomplcase} and \eqref{Effgeomcomplcase} the effective density and pressure are the following:
\begin{eqnarray}
\label{rhovcomplcase} 3H^2 &=& \rho_{\rm eff} = \frac{\rho}{f_R} + \frac{3\lambda}{2f_R}\rho - \frac{R}{2} + \frac{f}{2f_R} - 3H\frac{\dot{f}_R}{f_R}\;,\\
\label{pvcomplcase} -2\dot{H} - 3H^2 &=& p_{\rm eff} =  \frac{1}{f_R}\left(\ddot{f}_R + 2H\dot{f}_R\right) - \frac{f}{2f_R} - \frac{\lambda}{2f_R}\rho + \frac{R}{2}\;.
\end{eqnarray}
From Eq.~\eqref{rhovcomplcase} it appears that the energy density of the perfect fluid is rescaled by a factor $1/f_R$. Looking at Eq.~\eqref{Friede}, we can extract a form for $\rho_{\rm v}$ in the following way:
\begin{equation}\label{rhovcomplcase2} 
\rho_{\rm v} = \frac{\rho}{f_R} - \rho + \frac{3\lambda}{2f_R}\rho - \frac{R}{2} + \frac{f}{2f_R} - 3H\frac{\dot{f}_R}{f_R}\;,
\end{equation}
whereas the form of $p_{\rm v}$ is already given in Eq.~\eqref{pvcomplcase}, since our fluid is pressureless. From Eqs.~\eqref{pvcomplcase} and \eqref{rhovcomplcase2} we can write the following differential equation for $f_R$:
\begin{equation}\label{freqcompl}
 \ddot{f}_R - H \dot{f}_R - \left[\rho + \rho_{\rm v}\left(1 + w_{\rm v}\right)\right]f_R = -\rho(1 + \lambda)\;.
\end{equation}
Note, as a cross-check, that for $\rho_{\rm v} = \lambda = 0$ the above equation simplifies to
\begin{equation}\label{freqcompl2simpl}
 \ddot{f}_R - H\dot{f}_R - \rho f_R = - \rho\;,
\end{equation}
which possesses the particular solution $f_R = 1$, i.e. $f = R + \Lambda$, the original Einstein-Hilbert action plus an integration ``cosmological'' constant. We expect this solution to be the only one, otherwise there would exist an alternative $f(R)$ theory which would behave exactly as general relativity. Let us speculate a bit more on this point. If $\rho_{\rm v} = \lambda = 0$, i.e. for a pure Einstein-de Sitter universe, we have from Eq.~\eqref{Friede}
\begin{equation}
 H = \frac{2}{3t}\;, \qquad \rho = \rho_0\frac{t^2_0}{t^2}\;,
\end{equation}
where $t_0$ is the present cosmic time (i.e. the age of the universe). Considering the homogeneous part of Eq.~\eqref{freqcompl2simpl} and looking for a solution of the form $f_R \propto t^n$, we find:
\begin{equation}\label{freqcompl2simplhom}
 n(n - 1) - \frac{2}{3}n - \rho_0 t_0^2 = 0\;,
\end{equation}
which gives:
\begin{equation}\label{freqcompl2simplhomsol}
 n_{1,2} = \frac{5}{6} \pm \sqrt{\frac{25}{36} + \rho_0 t_0^2}\;,
\end{equation}
and the general solution of Eq.~\eqref{freqcompl2simpl} can be written as:
\begin{equation}\label{freqcompl2simplhomsol2}
 f_R = 1 + C_1\;t^{n_1} + C_2\;t^{n_2}\;.
\end{equation}
Now, the initial conditions we adopt here are $f_R(t_0) = 1$ and $\dot{f}_R(t_0) = 0$. The reason is essentially not spoiling the agreement between general relativity and solar system tests, see \cite{Capozziello:2005ku, Nojiri:2006gh, Nojiri:2006jy, Nojiri:2006be, Wu:2007tn}. However, we stress here that these two conditions also imply $C_1 = C_2 = 0$ and therefore restore the general relativity limit $f_R = 1$ when $\rho_{\rm v} = \lambda = 0$. 

Changing the variable to the redshift and employing Eqs.~\eqref{Friede} and \eqref{OmegavEvo} one can recast Eq.~\eqref{freqcompl} in the following compact form:
\begin{eqnarray}\label{freqcompl2}
 (1 + z)^2f''_R + \frac{1 + z}{2}\left[7 - \Omega_{\rm v}\left(1 + \frac{2\sqrt{\Omega_{\rm v}}}{b}\right)\right]f_R' - 3\left[1 - \Omega_{\rm v}\left(\frac{1}{3} + \frac{2\sqrt{\Omega_{\rm v}}}{3b}\right)\right]f_R = \nonumber\\ - 3(1 + \lambda)(1 - \Omega_{\rm v})\;,
\end{eqnarray}
where again the prime denotes derivation with respect to the redshift $z$. The curvature $R$ can be easily found as
\begin{equation}
 R = -6(\dot{H} + 2H^2) = -\frac{3H_0^2\Omega_{\rm m0}(1 + z)^3}{1 - \Omega_{\rm v}}\left[1 + \Omega_{\rm v}\left(1 + \frac{2\sqrt{\Omega_{\rm v}}}{b}\right)\right]\;.
\end{equation}
The initial conditions $f_R(t_0) = 1$ and $\dot{f}_R(t_0) = 0$, translated to the redshift variable, are
\begin{eqnarray}
 \left.\frac{d^2f}{dz^2}\right|_{z = 0} = \left.\frac{d^2R}{dz^2}\right|_{z = 0}\;, \qquad \left.\frac{df}{dz}\right|_{z = 0} = \left.\frac{dR}{dz}\right|_{z = 0}\;,\\
\end{eqnarray}
Finally, the initial condition on $f$ can be extracted by Eq.~\eqref{rhovcomplcase}, being that $\rho_{\rm v0} = 3H_0^2 - \rho_{\rm m0}$. Thus, we have
\begin{equation}
f(z = 0) = R(z = 0) + 6H_0^2\left(1 - \Omega_{\rm m0} - \frac{3}{2}\lambda\Omega_{\rm m0}\right)\;.
\end{equation}
Note the correction due to the $\lambda T$ term.
\begin{figure}[htbp]
  \includegraphics[width=0.45\columnwidth]{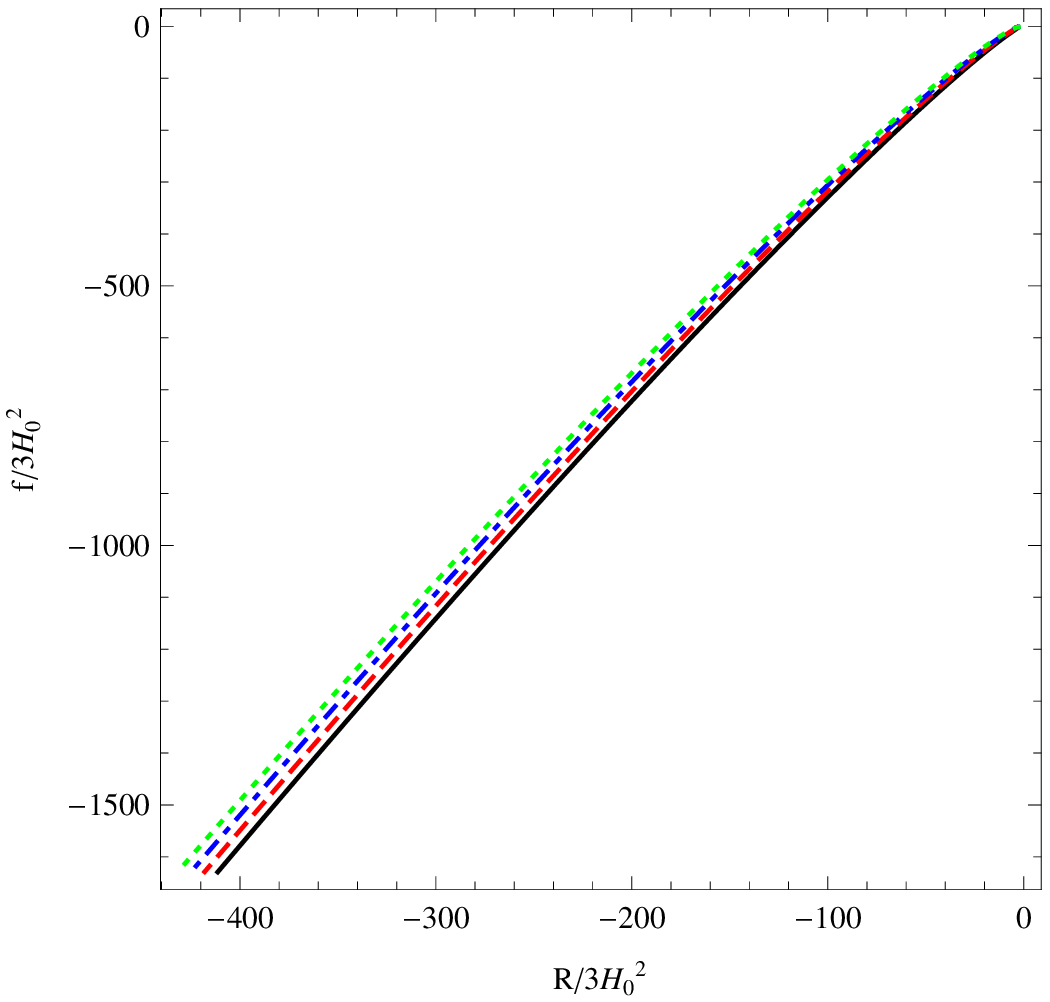}\;\includegraphics[width=0.45\columnwidth]{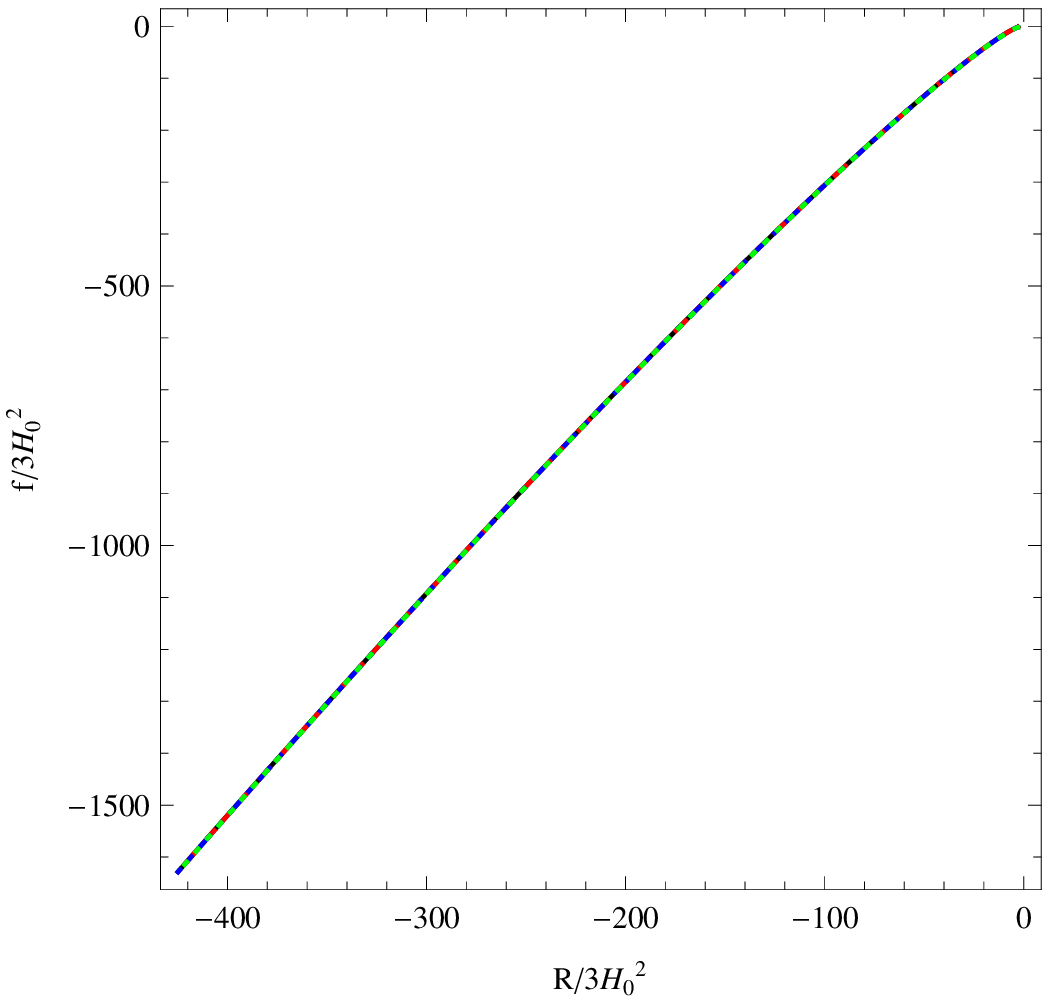}
\caption{Evolution of $f$ as a function of the curvature $R$. Left panel: $\lambda = 0$, i.e. we are considering a pure $f(R)$ theory, and $b = 0.6, 0.8, 1.0, 1.2$ (solid black, dashed red, dot-dashed blue and dotted green, respectively). Right panel: $b = 1.0$ and $\lambda = -0.2, 0, 0.2, 0.4$ (solid black, dashed red, dot-dashed blue and dotted green, respectively. The curves appear superposed.). We have chosen as initial conditions in $\Omega_{\rm m0} = 0.3$ the values $\Omega_{\rm v0} = 1 - \Omega_{\rm m0} = 0.7$. Note that $f$ and $R$ are normalised to $3H_0^2$ and the redshift interval chosen is $0 < z < 10$.}
\label{Fig3}
\end{figure}
In \figurename{ \ref{Fig3}} we plot the solution for $f$. Note that we normalise $f$ and $R$ to $3H_0^2$. In the left panel we fix $\lambda = 0$, i.e. we are actually considering a pure $f(R)$ theory, and vary $b$. In the right panel, on the other hand, we consider positive and negative values of $\lambda$. As one may note, $\lambda$ has a poor influence on the evolution of $f$. We could expect this from inspection of Eq.~\eqref{freqcompl2}. Indeed, $\lambda$ only enters the source term on the right hand side and therefore, when $\Omega_{\rm v}$ grows to unity, its impact on the evolution of $f$ is weak. On the other hand, larger values of $\lambda$ may have a relevant effect at early times, determining the slope of $f$.
\begin{figure}[ht]
  \includegraphics[width=0.45\columnwidth]{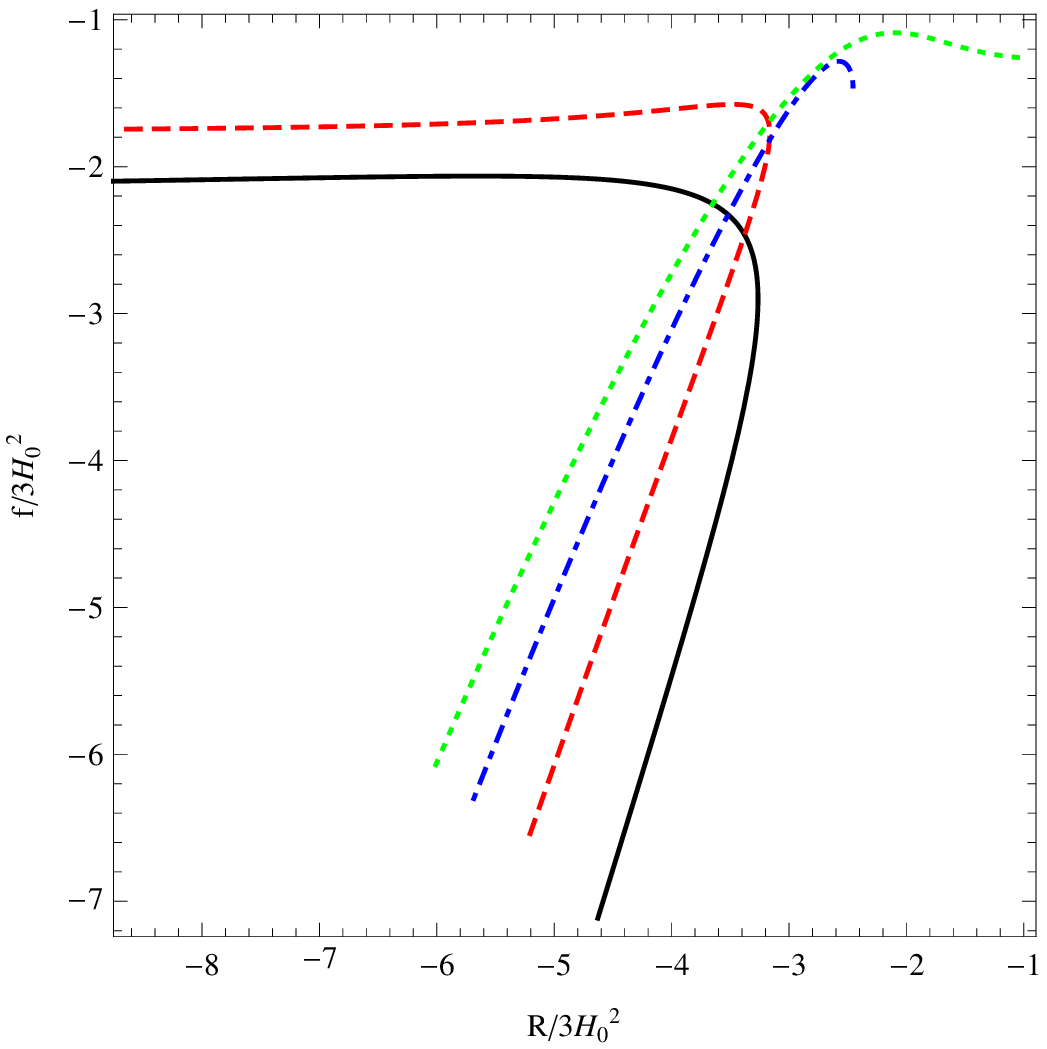}\;\includegraphics[width=0.45\columnwidth]{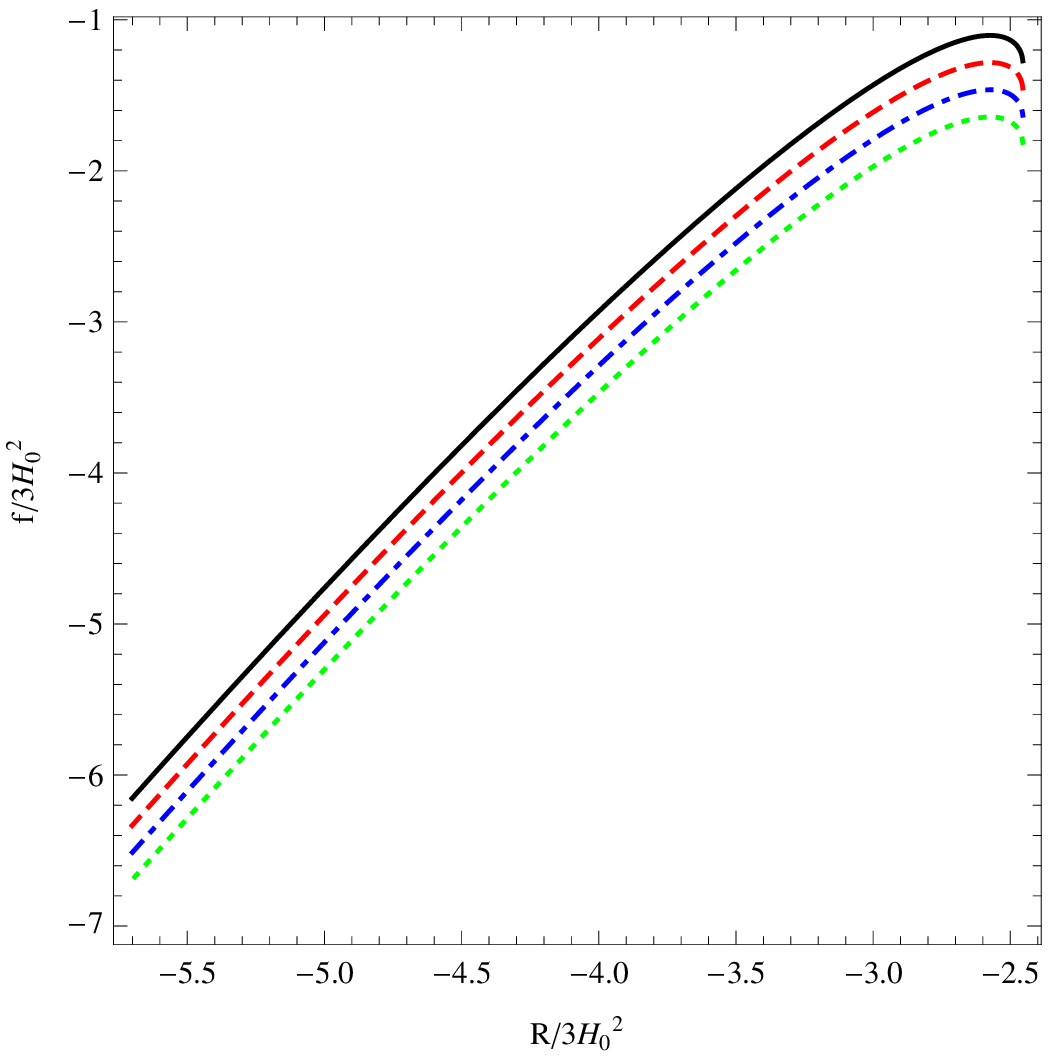}
\caption{Same as \figurename{ \ref{Fig3}}, but with $-0.9 < z < 1$. Note, in the left panel, that the evolution of $f$ starts from below for all the cases.}
\label{Fig4}
\end{figure}
\newpage
In \figurename{ \ref{Fig4}} and \figurename{ \ref{Fig5}} we display the future evolution of $f$, both as function of $R$ or of $z$. For the same reason stated above, the effect of $\lambda$ is not relevant.
\begin{figure}[ht]
  \includegraphics[width=0.45\columnwidth]{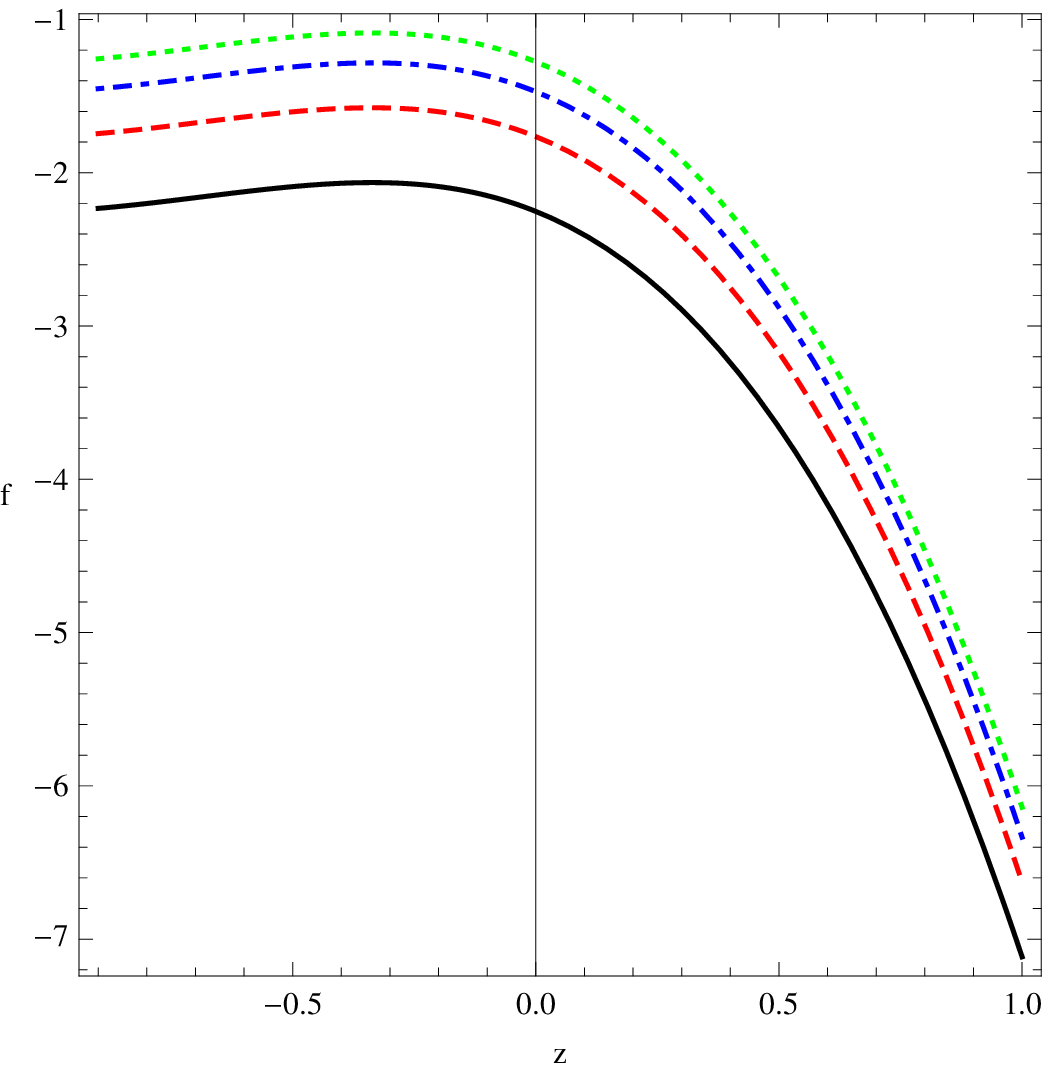}\;\includegraphics[width=0.45\columnwidth]{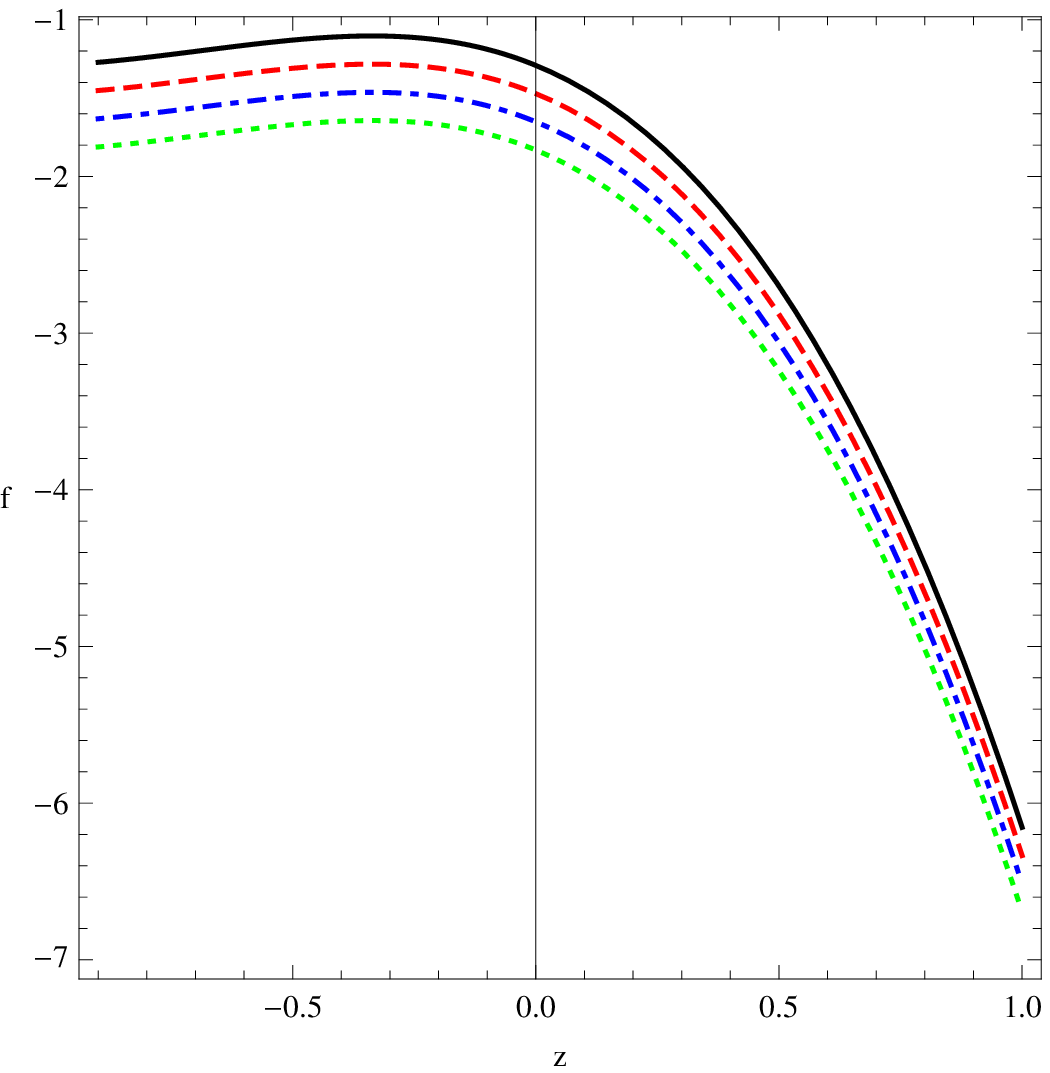}
\caption{Same as \figurename{ \ref{Fig4}}, but with $f$ as function of $z$.}
\label{Fig5}
\end{figure}
\newpage
Finally, in \figurename{ \ref{Fig6}} we plot the solution for $\Omega_{\rm v}$. Again, its evolution appears to be independent of $\lambda$.
\begin{figure}[ht]
  \includegraphics[width=0.45\columnwidth]{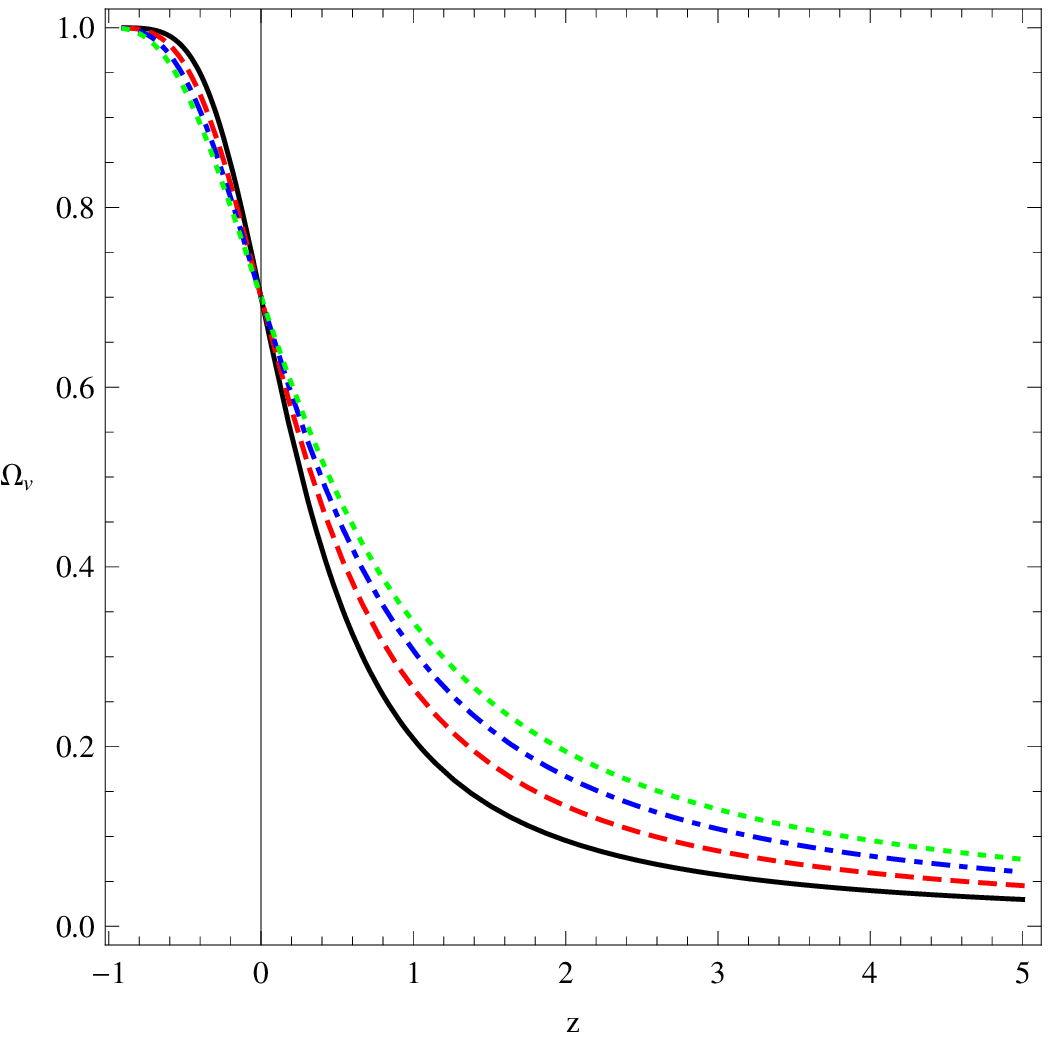}\;\includegraphics[width=0.45\columnwidth]{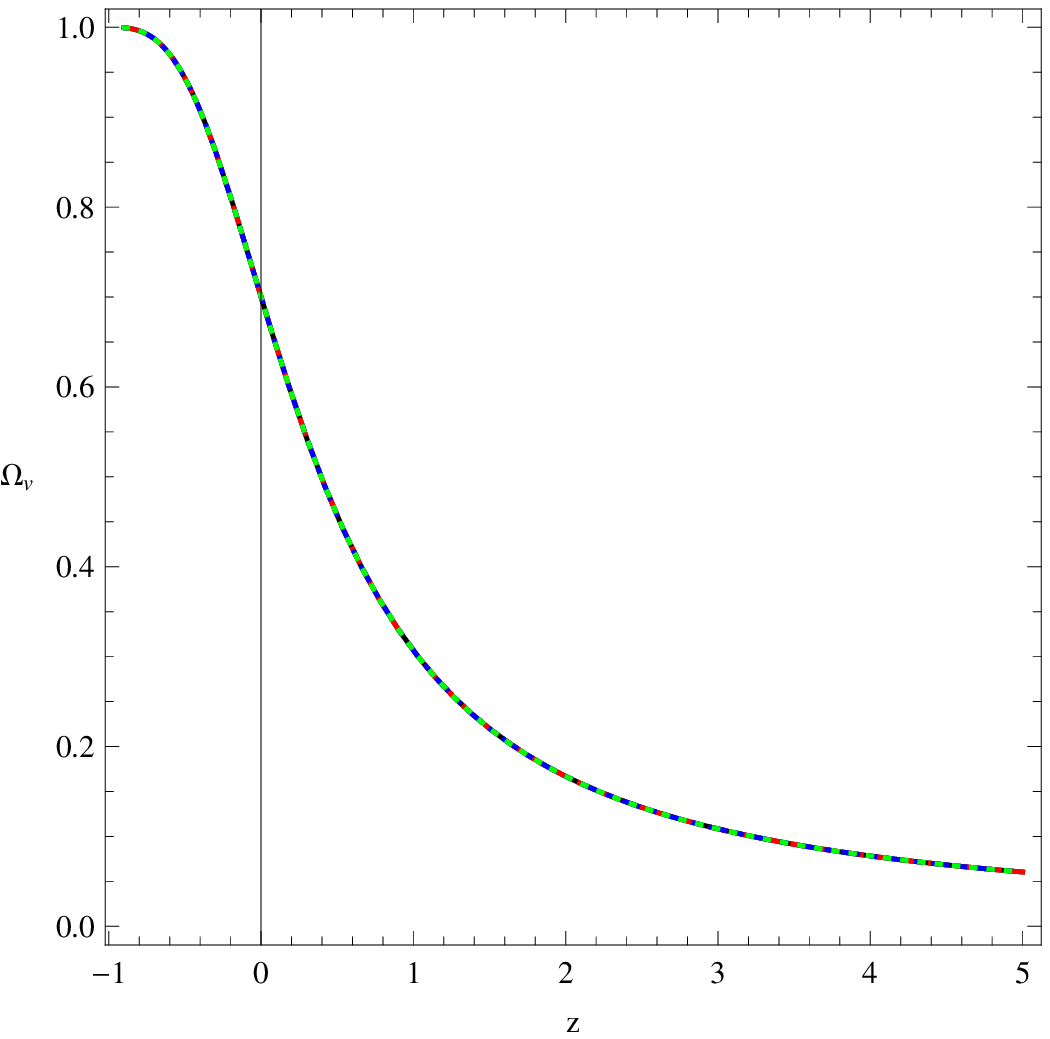}
\caption{Evolution of $\Omega_{\rm v}$ as a function of the redshift. Left panel: $\lambda = 0$, i.e. we are considering a pure $f(R)$ theory, and $b = 0.6, 0.8, 1.0, 1.2$ (black, dashed red, dash-dotted blue and dotted green, respectively). Right panel: $b = 1.0$ and $\lambda = -0.2, 0, 0.2, 0.4$ (black, dashed red, dash-dotted blue and dotted green, respectively). Note that the curves are superposed. We have chosen as initial conditions in $\Omega_{\rm m0} = 0.3$ the values $\Omega_{\rm v0} = 1 - \Omega_{\rm m0} = 0.7$.}
\label{Fig6}
\end{figure}

\newpage
%%%%%%%%%%%%%%%%%%%%%%%%%%%%%%%%%%%%%%%%%%%%%%%%%%%%%%%%%%%%%%%%%%%%%%%%%%%%%%%%%%%%%%%%%%%%%%%%%%%%%%%%%%%%%%%%

\section{Discussion and Conclusions}\label{Sec:DiscConcl}

In this work we have investigated a description of holographic dark energy in terms of suitably reconstructed $f(R, T)$ gravity theories. The latter have been recently introduced as modifications of Einstein's theory possessing some interesting solutions \textbf{which are} relevant in cosmology and astrophysics \cite{Harko:2011kv}.

We have \textbf{considered two special} types of models: $f(R,T) = R + 2f(T)$, i.e. a correction to the Einstein-Hilbert action depending on the matter content, and $f(R,T) = f(R) + \lambda T$, i.e. a simple $T$-linear correction to the class of $f(R)$ theories. 

Since we have assumed the matter content to be a pressureless perfect fluid, then $T = \rho$, i.e. the corrections assumed are directly dependent on the energy density of the universe content. We \textbf{have} constructed differential equations for the function $f$ under investigation and numerically solved \textbf{them}, physically specifying the required initial conditions. Our simple analysis shows that holographic dark energy models are contained in the larger class of $f(R,T)$ theories, at least considering a given background evolution of the universe. 

It would be interesting to investigate how the evolution of matter perturbations would change, depending on the description of dark energy. We expect, in principle, different results when using holographic dark energy or its $f(R,T)$ reconstruction and therefore there is possibility for discriminating between the two descriptions. For example, it would be interesting to adapt the recently proposed scheme for perturbations in $f(R)$ theories \cite{Bertacca:2011wu} to the broader $f(R,T)$ class. We leave this as a future work.

%%%%%%%%%%%%%%%%%%%%%%%%%%%%%%%%%%%%%%%%%%%%%%%%%%%%%%%%%%%%%%%%%%%%%%%%%%%%%%%%%%%%%%%%%%%%%%%%%%%%%%%%%%%%%%%%

\section*{Acknowledgements}

MJSH and OFP thank Professor S. D. Odintsov for useful comments and also CNPq (Brazil) for partial financial support. 

%%%%%%%%%%%%%%%%%%%%%%%%%%%%%%%%%%%%%%%%%%%%%%%%%%%%%%%%%%%%%%%%%%%%%%%%%%%%%%%%%%%%%%%%%%%%%%%%%%%%%%%%%%%%%%%%

\bibliographystyle{unsrt}

\end{document}